\documentclass[conference]{IEEEtran}
\IEEEoverridecommandlockouts
\usepackage{cite}
\usepackage{amsmath,amssymb,amsfonts}
\usepackage{algorithmic}
\usepackage{graphicx}
\usepackage{textcomp}
\usepackage{xcolor}
\usepackage{todonotes}
\def\BibTeX{{\rm B\kern-.05em{\sc i\kern-.025em b}\kern-.08em
    T\kern-.1667em\lower.7ex\hbox{E}\kern-.125emX}}
\begin{document}
\newcommand\copyrighttext{%
  \footnotesize This  work  has  been  submitted  to  the  IEEE  for  possible  publication.  Copyright  may  be  transferred  without  notice,  after  which  this  version  may  no  longer be accessible. 
  Conference: 2020 Ural Symposium on Biomedical Engineering, Radioelectronics and Information Technology (USBEREIT). Preprint available here: https://arxiv.org/abs/1912.04672}
\newcommand\copyrightnotice{%
\begin{tikzpicture}[remember picture,overlay]
\node[anchor=south,yshift=10pt] at (current page.south) {\fbox{\parbox{\dimexpr\textwidth-\fboxsep-\fboxrule\relax}{\copyrighttext}}};
\end{tikzpicture}%
}

\newpage

\title{Effects of lead position, cardiac rhythm variation and drug-induced QT prolongation on performance of machine learning methods for ECG processing\\
\thanks{The reported study wassupported by RFBR research project No. 19-37-50079 and supported by the IIF UrB RAS theme \#AAAA-A18-118020590031-8, RF Government Act \#211 of March 16, 2013, the Program of the Presidium RAS.}
}

\author{
\IEEEauthorblockN{Marat Bogdanov}
\IEEEauthorblockA{\textit{Department of Computational Mathematics and Cybernetics} \\
\textit{Ufa State Aviation Technical University}\\
Ufa, Russia\\
bogdanov\_marat@mail.ru}
\and
\IEEEauthorblockN{Salim Baigildin}
\IEEEauthorblockA{\textit{Faculty of Physics and Mathematics} \\
\textit{Bashkir State Pedagogical University}\\
Ufa, Russia\\
salim.sagadatovich@gmail.com}
\and
\IEEEauthorblockN{Aygul Fabarisova}
\IEEEauthorblockA{\textit{Institute of Natural Sciences and Mathematics} \\
\textit{Ural Federal University}\\
Ekaterinburg, Russia\\
aygul.fabarisova@gmail.com}
\and
\IEEEauthorblockN{Konstantin Ushenin}
\IEEEauthorblockA{\textit{Institute of Natural Sciences and Mathematics} \\
\textit{Ural Federal University}\\
Ekaterinburg, Russia \\
konstantin.ushenin@urfu.ru}
\and
\IEEEauthorblockN{Olga Solovyova}
\IEEEauthorblockA{\textit{Laboratory of Mathematical Physiology} \\
\textit{Institute of Immunology and Physiology UrB RAS}\\
Ekaterinburg, Russia \\
o.solovyova@iip.uran.ru}
}
\maketitle
\copyrightnotice
\begin{abstract}
Machine learning shows great performance in various problems of electrocardiography (ECG) signal analysis. However, collecting a dataset for biomedical engineering is a very difficult task. Any dataset for ECG processing contains from 100 to 10,000 times fewer cases than datasets for image or text analysis. This issue is especially important because of physiological phenomena that can significantly change the morphology of heartbeats in ECG signals. In this preliminary study, we analyze the effects of lead choice from the standard ECG recordings, variation of ECG during 24-hours, and the effects of QT-prolongation agents on the performance of machine learning methods for ECG processing. We choose the problem of subject identification for analysis, because this problem may be solved for almost any available dataset of ECG data. In a discussion, we compare our findings with observations from other works that use machine learning for ECG processing with different problem statements. Our results show the importance of training dataset enrichment with ECG signals acquired in specific physiological conditions for obtaining good performance of ECG processing for real applications.
\end{abstract}


\section{Introduction}

Machine learning is a widely used approach for the processing of physiological signals. This group of methods shows high performance in various problems: classification of electrocardiography (ECG) signals \cite{clifford2017af,kiranyaz2015real}, segmentation of ECG signals \cite{sereda2019ecg}, subject identification \cite{fratini2015individual,pinto2018evolution,lugovaya2005biometric}, and many other. An ECG processing also frequently complements processing of other physiological signals \cite{vasilyev2019case,kublanov2017classification}.

However, machine learning methods require a training dataset, and usually, databases for biomedical engineering are significantly smaller than for other applications. As an illustration, MNIST is a basic dataset for the problem of handwritten digits recognition. It contains 60,000 training images and 10,000 testing images. The biggest open dataset in PhysioNet \cite{goldberger2000physiobank} is the PTB Diagnostic ECG Database \cite{bousseljot1995nutzung} that contains only 549 recordings from 294 subjects.

The small size of the training dataset raises a question about the effect of physiological variations and methods of ECG registration on the performance of proposed solutions. In our current research, we were aimed to analyze the effect of human cardiac physiology phenomena on the performance of machine learning algorithms for ECG processing. We have chosen \textit{the subject identification problem} for our goal. This problem statement is similar to fingerprint identification, but it uses ECG as input information. 

A solution of the problem is usually based on the methods that extract features from ECG and methods that classify the extracted features to classes, where each class is a unique subject. Also, a major part of the proposed solutions requires a short ECG fragment with length from one cardiac cycle to 5-minute records. For this reason, the subject identification problem may be solved and analyzed with almost any available dataset of ECG data. This is a great advantage of this problem for our goals over the problem of ECG classification, PQRST complex segmentation, and many others. A wide description of subject identification methods may be found in recent reviews \cite{fratini2015individual,pinto2018evolution}.

In this preliminary study, we used a solution that was proposed in the previous work of the co-authors \cite{bogdanov2018factors,bogdanov2018statistical}. That solution used the morphology of PQRST complexes and indirectly included information about the heart rate variability. Here, we test this solution against the choice of ECG lead position, long-time variability of heart rate, and the effects of QT-prolongation drugs (class III). In the discussion, we compare our results with other observations and show how our findings may be helpful to a wide area of studies that use machine learning approaches to ECG analysis.

\section{Methods}

\subsection{Databases}

\textit{Physikalisch-Technische Bundesanstalt Diagnostic ECG Database (PTB Database)} \cite{bousseljot1995nutzung,goldberger2000physiobank} was chosen for analysis of the effect of a lead choice on the subject identification problem. This database includes 549 records from 290 subjects. Each subject is represented by one to five records. Each record includes the conventional 12 leads (I, II, III, aVR, aVL, avf, V1, V2, V3, V4, V5, V6), and the three Frank leads (Vx, Vy, Vz).
 
\textit{The Long Term ST Database (LTSTD)} \cite{jager2003long,goldberger2000physiobank} was chosen for analysis of the effect of 24-hours rhythm variations to the subject identification problem. The LTSTD contains 86 lengthy ECG recordings of 80 human subjects, chosen to exhibit a variety of events of ST-segment changes, including ischemic ST episodes, axis-related non-ischemic ST episodes, episodes of slow ST level drift, and episodes containing mixtures of these phenomena.

In addition to ischemic events, any long-term ECG recording includes slight variations of the PQRST complex caused by the regulation of parasympathetic and sympathetic nervous systems, which is related to a person's diurnal cycles and some stress surroundings.

\textit{The ECG Effects of Ranolazine, Dofetilide, Verapamil, and Quinidine database (ECGRDVQ database)} \cite{johannesen2014differentiating,goldberger2000physiobank} was chosen for analysis of QT-prolonging drugs on the person identification problem. That database contains ECG recordings of 22 healthy subjects for 24 hours under the effect of dofetilide (500 $\mu$g), quinidine sulfate (400 mg), ranolazine (1500 mg), verapamil hydrochloride (120 mg). QT-prolonging drugs affect on duration of the transmembrane action potential of cardiomyocytes, which causes changes in the T-wave shape and QT interval prolongation.

\begin{table*}[t]
\begin{tabular}{l|llllllllllll|rr}
Method & I    & II    & III    & aVR    & aVL    & aVF    & V1    & V2    & V3    & V4   & V5   & V6 & MIN   & MAX-MIN \\ \hline
multi-layer perceptron      & 97\% & \textbf{98\%} & 96\% & 96\% & 97\% & 96\% & 97\% & 97\% & \textbf{98\%} & \textbf{98\%} & 97\% & \textbf{98\%} & 96\% & 2\%       \\
naive Bayes classifier     & 54\% & 45\% & 46\% & 41\% & 46\% & 45\% & 53\% & \textbf{60\%} & \textbf{60\%} & 58\% & 52\% & 49\% & 41\% & 19\%      \\
decision tree classifier      & 69\% & 70\% & 67\% & 69\% & 70\% & 69\% & 72\% & 73\% & \textbf{76\%} & \textbf{76\%} & 70\% & 74\% & 69\% & 9\%       \\
extra-trees classifier      & 96\% & \textbf{97\%} & 92\% & 93\% & 94\% & 93\% & 96\% & \textbf{97\%} & \textbf{97\%} & \textbf{97\%} & 96\% & 96\% & 92\% & 5\%       \\
k-nearest neighbour votes      & 92\% & 91\% & 89\% & 86\% & 93\% & 89\% & 93\% & \textbf{95\%} & 93\% & 94\% & 92\% & \textbf{95\%} & 86\% & 9\%       \\
linear discriminant analysis      & \textbf{92\%} & 91\% & 82\% & 82\% & 89\% & 87\% & 86\% & 87\% & 86\% & 86\% & 88\% & 88\% & 82\% & 10\%      \\
linear support vector classifier      & 82\% & 85\% & 83\% & 82\% & 82\% & 84\% & 89\% & 90\% & \textbf{93\%} & 90\% & 84\% & 83\% & 82\% & 11\%      \\
logistic regression classifier      & 95\% & 97\% & 94\% & 92\% & 96\% & 95\% & 96\% & 97\% & 97\% & \textbf{98\%} & 96\% & 96\% & 92\% & 6\%       \\
nearest centroid classifier      & 62\% & 53\% & 51\% & 50\% & 53\% & 53\% & 57\% & \textbf{63\%} & 61\% & 61\% & 56\% & 59\% & 50\% & 13\%      \\
random forest classifier     & 10\% & 10\% & 8\%  & 13\% & \textbf{14\%} & 7\%  & 9\%  & 7\%  & 6\%  & 7\%  & 10\% & 10\% & 6\% & 8\%       \\
ridge regression classifier     & 39\% & \textbf{46\%} & 37\% & 36\% & 44\% & 39\% & 38\% & 43\% & 43\% & 39\% & 38\% & 38\% & 36\% & 10\%      \\
Gaussian mixture model     & 63\% & 52\% & 53\% & 52\% & 57\% & 54\% & 66\% & \textbf{73\%} & 72\% & 67\% & 60\% & 58\% & 52\% & 21\%      \\
support vector machine     & \textbf{70\%} & 66\% & 57\% & 63\% & 62\% & 59\% & 65\% & 66\% & 67\% & 67\% & 65\% & 64\% & 59\% & 13\%     
\end{tabular}\caption{The dependency between ECG lead and the performance of the subject identification solutions. The last two columns show the minimal accuracy and the difference between the maximum and minimum ones for each method.}\label{tab:leads}
\end{table*}

\begin{figure}[h] 
	\includegraphics[width=0.49\textwidth]{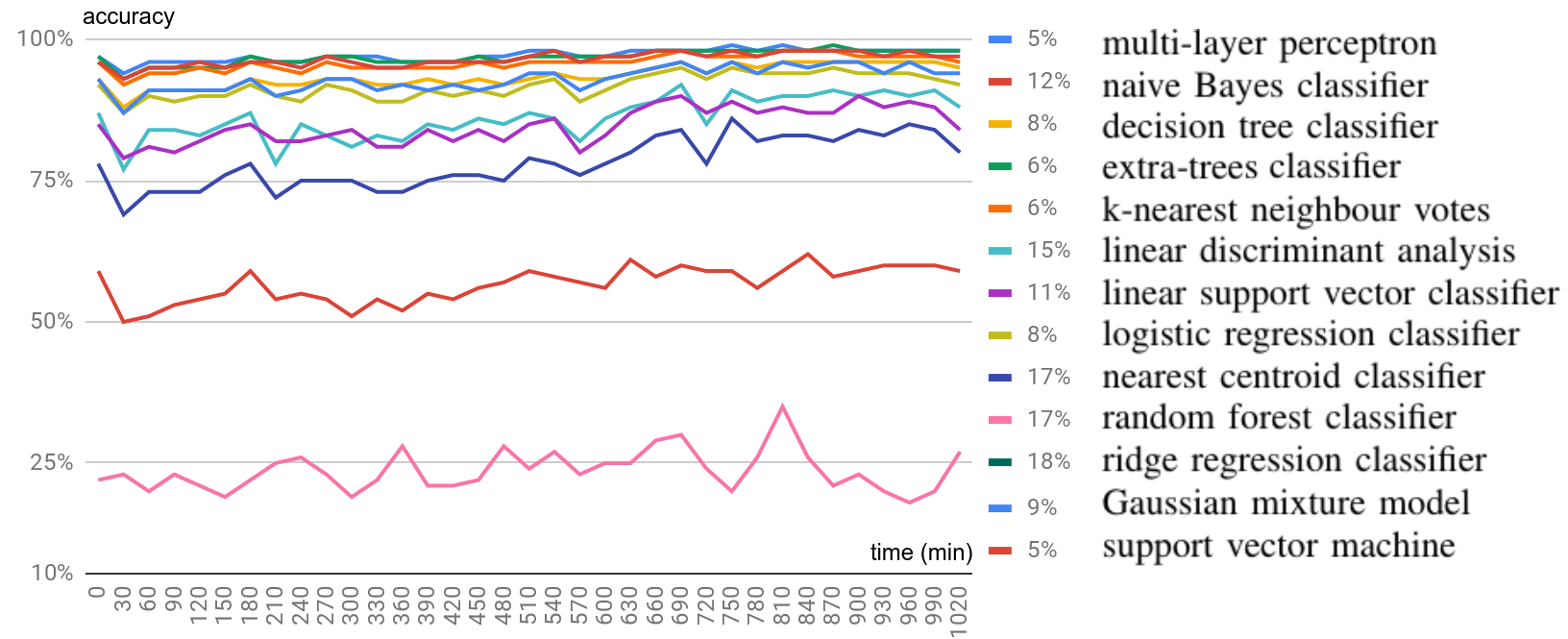}
	\caption{Changes in the subject identification rate (classification accuracy) with accounting for the effects of normal physiological variation in the human ECG over 24 hours. Percent near to plots shows a difference between the maximal and minimal accuracy.}\label{fig:time}
\end{figure}

\begin{table*}[h]
\begin{tabular}{l|ll|l|l}
                                                               & \begin{tabular}[c]{@{}l@{}}Train set: before drugs\\ Validation set: before drugs\end{tabular} & \begin{tabular}[c]{@{}l@{}}Train set: before drugs\\ Validation set: after drugs\end{tabular} & \begin{tabular}[c]{@{}l@{}}Accuracy \\ reduction\end{tabular} & \begin{tabular}[c]{@{}l@{}}Train set: before+afrer drugs\\ Validation set: after drugs\end{tabular} \\ \hline
multi-layer perceptron                                                            & 98\%                                                                                     & 88\%                                                                                    & 10\%                                                        & 100\%                                                                                         \\
naive Bayes classifier & 79\%                                                                                     & 63\%                                                                                    & 16\%                                                        & 90\%                                                                                          \\
decision tree classifier                                  & 91\%                                                                                     & 52\%                                                                                    & 39\%                                                        & 93\%                                                                                          \\
extra-trees classifier                    & 94\%                                                                                     & 85\%                                                                                    & 9\%                                                         & 100\%                                                                                         \\
k-nearest neighbour votes     & 97\%                                                                                     & 90\%                                                                                    & 7\%                                                         & 96\%                                                                                          \\
linear discriminant analysis                                 & 95\%                                                                                     & 90\%                                                                                    & 5\%                                                         & 97\%                                                                                          \\
linear support vector classifier& 90\%                                                                                     & 80\%                                                                                    & 10\%                                                        & 97\%                                                                                          \\
logistic regression classifier           & 96\%                                                                                     & 80\%                                                                                    & 16\%                                                        & 98\%                                                                                          \\
nearest centroid classifier                                  & 81\%                                                                                     & 83\%                                                                                    & -2\%                                                        & 89\%                                                                                          \\
random forest classifier                                  & 47\%                                                                                     & 41\%                                                                                    & 6\%                                                         & 67\%                                                                                          \\
ridge regression classifier                           & 72\%                                                                                     & 69\%                                                                                    & 3\%                                                         & 74\%                                                                                          \\
Gaussian mixture model                                      & 95\%                                                                                     & 90\%                                                                                    & 5\%                                                         & 97\%                                                                                          \\
support vector machine                                                            & 70\%                                                                                     & 90\%                                                                                    & -20\%                                                       & 95\%                                                                                         
\end{tabular}\caption{Reduction of subject identification rate (classification accuracy) under the effect of QT-prolongation agents.}\label{tab:drugs}
\end{table*}

\subsection{ECG processing}

In the current study, the subject identification problem is considered as a classification problem that should be solved with machine learning approaches. Each ECG signal is represented as a vector that was used as input of a classification algorithm, and each subject (individual) corresponds to the target class of the algorithm output. In this case, the identification ratio is equivalent to classification accuracy. The subject identification problem was established for the short ECG recorded from a single lead with a length of 20 heartbeats.

The processing procedure was as follows. First, the whole signal was separated in individual heartbeat signals. Then, the heartbeat signals were aligned to the peak of the R wave. The position of each peak in time and its amplitude provided nine features for a heartbeat. Features that were extracted from all heartbeats of the short fragment were joined together for the creation of a long input vector of features with 180 components.

The performance analysis was performed for each database independently. The training dataset was formed from the vectors corresponding to the first 20 heartbeats of the first record of each patient. Validation datasets were formed in various ways. For the PTB database, the only one validation dataset was formed from randomly chosen ECG fragments of the normal rhythm for each patient. For the LTSTB and ECGRDVQ databases, several validation datasets were formed for each half-hour of recordings. These datasets include the first 20 complexes at the beginning of each half-hour intervals of long-time monitored ECG.

The 14 algorithms from the \textit{scikit-learn} package \cite{pedregosa2011scikit} were used for classification: a multi-layer perceptron, naive Bayes classifier for multivariate Bernoulli distributions, a decision tree classifier, an extra-trees classifier, k-nearest neighbor votes, a linear discriminant analysis, a linear support vector classifier, a logistic regression classifier, a nearest centroid classifier, a random forest classifier, a ridge regression classifier, a ridge classifier with built-in cross-validation, a Gaussian mixture model, and a support vector machine.  The extended description and analysis of the subject identification problem also can be found in the previous work of the co-authors \cite{bogdanov2018factors,bogdanov2018statistical}.

\section{Results}

Table \ref{tab:leads} shows the analysis of subject identification accuracy for each of the 12 conventional ECG leads. The presented table shows that methods with high accuracy are almost independent of the lead choice for subject identification. This proposition is confirmed with non-parametric rank correlation coefficients: corr.=-0.52 Spearman's R ($p < 0.056$), corr.=-0.411 Kendall's $\tau$ (0.046). The best accuracy was obtained using three methods: multi-layer perception, extremely randomized tree classifier, and logistic regression. These methods also provide the smallest difference between the minimal and maximal accuracy values across all leads. Thus, we should assume that the performance of machine learning approaches is almost independent of chosen ECG leads.

Figure \ref{fig:time} presents an analysis of subject identification accuracy under normal 24-hour rhythm variability. The series of validation datasets is built-on short ECG fragments that are spaced through 30-minute intervals on the timeline. The minimal variations in accuracy observed for the multi-layer perceptron, extremely randomized tree, and support vector machines.

Accordingly, the structure of plots, we conclude that algorithms with high accuracy are almost independent of the variability of ECG during the 24-hour, long-time monitoring period.

Table \ref{tab:drugs} shows the effect of QT-prolongation drugs on the problem of subject identification by ECG. This effect cannot be ignored in a practical application because accuracy reduction may reach 40\%. However, the accuracy of the algorithm is recovered if the training dataset is extended with ECG signals after taking of the drug.

We observe weak correlation between the performance of the algorithm on normal conditions and performance of the algorithm after taking the drugs if this algorithm does not have data with prolonged QT interval in the training dataset (corr.=0.52, $p < 0.051$, Spearman's R; corr.=0.40, p<0.05m Kendall's $\tau$). However, machine learning approaches with high performance on normal data, show high performance on both types of data if the training dataset is extended with additional cases (corr.=0.80, $p < 0.0005$, Spearman's R; corr.=0.64, $p<0.0019$, Kendall's $\tau$).

\section{Discussion}

Works \cite{hoekema2001geometrical,schijvenaars2000intra} show a significant difference in morphological parameters of heartbeats in ECG signals from different leads. However, our study shows that algorithm performance is almost independent of the lead choice. This observation is consistent with other works that are devoted to the subject identification problem \cite{fratini2015individual} and clinical applications of ECG analysis \cite{hannun2019cardiologist,clifford2017af,kiranyaz2015real}.

We observed the reduction of the subject identification ratio caused by changes in ECG over time. This type of ECG variations is usually named as an intra-subject ECG variation in time. Previous work studied these phenomena and showed the reduction of subject identification ratio because of the following reasons:
changes in ECG over the time \cite{sansone2013influence} (7\% reduction in 120 minute monitoring); position, exercises and clinical stress-test \cite{poree2016ecg,wahabi2014evaluating, wang2014research,sidek2014ecg,sidek2011automobile}. However, some classification algorithms used in our study show a less than 5\% reduction in the performance. This result is smaller than that presented in other works \cite{sansone2013influence,poree2016ecg}. This could be explained by either the more effective approach taken here, or the absence of physical stress in patients undergoing a long-time ECG monitoring for LTSTD.

We observe significant negative effects of QT-prolongation drugs on the subject-identification ratio. Effect of drugs on the subject identification problem is not mentioned in reviews of 2015 \cite{fratini2015individual} and 2019 \cite{pinto2018evolution} years. 

In summary, we can speculate that a wide area of problem statements for ECG processing may be solved by using a signal from the only one ECG lead. However, the performance of any algorithm of ECG processing may be reduced because of ECG variability in 24-hours, ECG changes in physical stress conditions, and QT-prolongation agents. According to our results and earlier reports \cite{poree2016ecg,wahabi2014evaluating}, compensation of this reduction requires enrichment of the training dataset with ECG from a subject exposed to the conditions that affect cardiac electrophysiology.

\section{Conclusion}

In this study, we have analyzed the performance of the machine learning approaches for the subject identification from the ECG signals. Some results may be extended to any method of ECG processing that uses the machine learning approach. Based on our results and literature observations, we suppose that the methods are notably sensitive to the following physiological phenomena: ECG variability in 24-hours, ECG changes in the stress condition, and QT segment prolongation due to effects of class III antiarrhythmic agents (the strongest effect). The reduction in the machine learning performance in real applications may be overcome by the enrichment of the training dataset with ECG acquired when the subjects are exposed to physical load or take medications (particularly, QT-prolongation agents) during ECG evaluation.

\bibliographystyle{IEEEtran}
\bibliography{bibliography}

\begin{thebibliography}{10}
\providecommand{\url}[1]{#1}
\csname url@samestyle\endcsname
\providecommand{\newblock}{\relax}
\providecommand{\bibinfo}[2]{#2}
\providecommand{\BIBentrySTDinterwordspacing}{\spaceskip=0pt\relax}
\providecommand{\BIBentryALTinterwordstretchfactor}{4}
\providecommand{\BIBentryALTinterwordspacing}{\spaceskip=\fontdimen2\font plus
\BIBentryALTinterwordstretchfactor\fontdimen3\font minus
  \fontdimen4\font\relax}
\providecommand{\BIBforeignlanguage}[2]{{%
\expandafter\ifx\csname l@#1\endcsname\relax
\typeout{** WARNING: IEEEtran.bst: No hyphenation pattern has been}%
\typeout{** loaded for the language `#1'. Using the pattern for}%
\typeout{** the default language instead.}%
\else
\language=\csname l@#1\endcsname
\fi
#2}}
\providecommand{\BIBdecl}{\relax}
\BIBdecl

\bibitem{clifford2017af}
G.~D. Clifford, C.~Liu, B.~Moody, H.~L. Li-wei, I.~Silva, Q.~Li, A.~Johnson,
  and R.~G. Mark, ``Af classification from a short single lead ecg recording:
  the physionet/computing in cardiology challenge 2017,'' in \emph{2017
  Computing in Cardiology (CinC)}.\hskip 1em plus 0.5em minus 0.4em\relax IEEE,
  2017, pp. 1--4.

\bibitem{kiranyaz2015real}
S.~Kiranyaz, T.~Ince, and M.~Gabbouj, ``Real-time patient-specific ecg
  classification by 1-d convolutional neural networks,'' \emph{IEEE
  Transactions on Biomedical Engineering}, vol.~63, no.~3, pp. 664--675, 2015.

\bibitem{sereda2019ecg}
I.~Sereda, S.~Alekseev, A.~Koneva, R.~Kataev, and G.~Osipov, ``Ecg segmentation
  by neural networks: errors and correction,'' in \emph{2019 International
  Joint Conference on Neural Networks (IJCNN)}.\hskip 1em plus 0.5em minus
  0.4em\relax IEEE, 2019, pp. 1--7.

\bibitem{fratini2015individual}
A.~Fratini, M.~Sansone, P.~Bifulco, and M.~Cesarelli, ``Individual
  identification via electrocardiogram analysis,'' \emph{Biomedical engineering
  online}, vol.~14, no.~1, p.~78, 2015.

\bibitem{pinto2018evolution}
J.~R. Pinto, J.~S. Cardoso, and A.~Louren{\c{c}}o, ``Evolution, current
  challenges, and future possibilities in ecg biometrics,'' \emph{IEEE Access},
  vol.~6, pp. 34\,746--34\,776, 2018.

\bibitem{lugovaya2005biometric}
T.~Lugovaya, ``Biometric human identification based on
  electrocardiogram.[master's thesis] electrotechnical university “leti”,''
  \emph{Saint-Petersburg, Russian Federation}, 2005.

\bibitem{vasilyev2019case}
V.~Vasilyev, V.~Borisov, A.~Syskov, and V.~Kublanov, ``Case study of
  interrelation between brain-computer interface based multimodal metric and
  heart rate variability,'' in \emph{12th International Conference on Health
  Informatics, HEALTHINF 2019-Part of 12th International Joint Conference on
  Biomedical Engineering Systems and Technologies, BIOSTEC 2019}.\hskip 1em
  plus 0.5em minus 0.4em\relax SciTePress, 2019, pp. 532--538.

\bibitem{kublanov2017classification}
V.~Kublanov, D.~Yamaliev, A.~Dolganov, and E.~Goncharova, ``Classification of
  the physical training level by heart rate variability and stabilography
  data,'' in \emph{2017 Siberian Symposium on Data Science and Engineering
  (SSDSE)}.\hskip 1em plus 0.5em minus 0.4em\relax IEEE, 2017, pp. 49--54.

\bibitem{goldberger2000physiobank}
A.~L. Goldberger, L.~A. Amaral, L.~Glass, J.~M. Hausdorff, P.~C. Ivanov, R.~G.
  Mark, J.~E. Mietus, G.~B. Moody, C.-K. Peng, and H.~E. Stanley, ``Physiobank,
  physiotoolkit, and physionet: components of a new research resource for
  complex physiologic signals,'' \emph{Circulation}, vol. 101, no.~23, pp.
  e215--e220, 2000.

\bibitem{bousseljot1995nutzung}
R.~Bousseljot, D.~Kreiseler, and A.~Schnabel, ``Nutzung der ekg-signaldatenbank
  cardiodat der ptb {\"u}ber das internet,'' \emph{Biomedizinische
  Technik/Biomedical Engineering}, vol.~40, no.~s1, pp. 317--318, 1995.

\bibitem{bogdanov2018factors}
M.~Bogdanov, V.~Kartak, A.~Dumchikov, and A.~Fabarisova, ``Factors influencing
  accuracy of biometrical personal identification based on cardiograms,''
  \emph{Pattern Recognition and Image Analysis}, vol.~28, no.~3, pp. 421--426,
  2018.

\bibitem{bogdanov2018statistical}
------, ``Statistical assessment of biometrical signs by electrocardiography,''
  \emph{Russian Journal of Cardiology}, vol.~23, no.~7, pp. 84--91, 2018.

\bibitem{jager2003long}
F.~Jager, A.~Taddei, G.~B. Moody, M.~Emdin, G.~Antoli{\v{c}}, R.~Dorn,
  A.~Smrdel, C.~Marchesi, and R.~G. Mark, ``Long-term st database: a reference
  for the development and evaluation of automated ischaemia detectors and for
  the study of the dynamics of myocardial ischaemia,'' \emph{Medical and
  Biological Engineering and Computing}, vol.~41, no.~2, pp. 172--182, 2003.

\bibitem{johannesen2014differentiating}
L.~Johannesen, J.~Vicente, J.~Mason, C.~Sanabria, K.~Waite-Labott, M.~Hong,
  P.~Guo, J.~Lin, J.~S. S{\o}rensen, L.~Galeotti \emph{et~al.},
  ``Differentiating drug-induced multichannel block on the electrocardiogram:
  randomized study of dofetilide, quinidine, ranolazine, and verapamil,''
  \emph{Clinical Pharmacology \& Therapeutics}, vol.~96, no.~5, pp. 549--558,
  2014.

\bibitem{pedregosa2011scikit}
F.~Pedregosa, G.~Varoquaux, A.~Gramfort, V.~Michel, B.~Thirion, O.~Grisel,
  M.~Blondel, P.~Prettenhofer, R.~Weiss, V.~Dubourg \emph{et~al.},
  ``Scikit-learn: Machine learning in python,'' \emph{Journal of machine
  learning research}, vol.~12, no. Oct, pp. 2825--2830, 2011.

\bibitem{hoekema2001geometrical}
R.~Hoekema, G.~J. Uijen, and A.~Van~Oosterom, ``Geometrical aspects of the
  interindividual variability of multilead ecg recordings,'' \emph{IEEE
  Transactions on Biomedical Engineering}, vol.~48, no.~5, pp. 551--559, 2001.

\bibitem{schijvenaars2000intra}
R.~Schijvenaars, ``Intra-individual variability of the electrocardiogram,''
  \emph{Assessment and exploitation in computerized ECG analysis [Ph. D.
  thesis]}, 2000.

\bibitem{hannun2019cardiologist}
A.~Y. Hannun, P.~Rajpurkar, M.~Haghpanahi, G.~H. Tison, C.~Bourn, M.~P.
  Turakhia, and A.~Y. Ng, ``Cardiologist-level arrhythmia detection and
  classification in ambulatory electrocardiograms using a deep neural
  network,'' \emph{Nature medicine}, vol.~25, no.~1, p.~65, 2019.

\bibitem{sansone2013influence}
M.~Sansone, A.~Fratini, M.~Cesarelli, P.~Bifulco, A.~Pepino, M.~Romano,
  F.~Gargiulo, and C.~Sansone, ``Influence of qt correction on temporal and
  amplitude features for human identification via ecg,'' in \emph{2013 IEEE
  Workshop on Biometric Measurements and Systems for Security and Medical
  Applications}.\hskip 1em plus 0.5em minus 0.4em\relax IEEE, 2013, pp. 22--27.

\bibitem{poree2016ecg}
F.~Por{\'e}e, G.~Kervio, and G.~Carrault, ``Ecg biometric analysis in different
  physiological recording conditions,'' \emph{Signal, image and video
  processing}, vol.~10, no.~2, pp. 267--276, 2016.

\bibitem{wahabi2014evaluating}
S.~Wahabi, S.~Pouryayevali, S.~Hari, and D.~Hatzinakos, ``On evaluating ecg
  biometric systems: session-dependence and body posture,'' \emph{IEEE
  Transactions on Information Forensics and Security}, vol.~9, no.~11, pp.
  2002--2013, 2014.

\bibitem{wang2014research}
Z.~Wang and Y.~Zhang, ``Research on ecg biometric in cardiac irregularity
  conditions,'' in \emph{2014 International Conference on Medical
  Biometrics}.\hskip 1em plus 0.5em minus 0.4em\relax IEEE, 2014, pp. 157--163.

\bibitem{sidek2014ecg}
K.~A. Sidek, I.~Khalil, and H.~F. Jelinek, ``Ecg biometric with abnormal
  cardiac conditions in remote monitoring system,'' \emph{IEEE Transactions on
  systems, man, and cybernetics: systems}, vol.~44, no.~11, pp. 1498--1509,
  2014.

\bibitem{sidek2011automobile}
K.~A. Sidek and I.~Khalil, ``Automobile driver recognition under different
  physiological conditions using the electrocardiogram,'' in \emph{2011
  Computing in Cardiology}.\hskip 1em plus 0.5em minus 0.4em\relax IEEE, 2011,
  pp. 753--756.

\end{thebibliography}
\end{document}